\documentstyle[aps,preprint]{revtex}
\draft
\def\gsim
{\raise2pt\hbox
{$\displaystyle{}\mathrel{\mathop>_{\raise1pt\hbox{$\sim$}}}{}$}} 
\def\lsim {\raise2pt\hbox
{$\displaystyle{}\mathrel{\mathop<_{\raise1pt\hbox{$\sim$}}}{}$}} %
\begin{document}
\title{Dissipation Enhanced Asymmetric Transport in Quantum Ratchets} 
\author{Gen Tatara, Macoto Kikuchi$^{1}$, Satoshi Yukawa$^{2}$
and Hiroshi Matsukawa$^{1}$}
\address{Department of Earth and Space Science, Osaka University, Toyonaka, 
Osaka 560\\ 
$^{1}$ Department of Physics, Osaka University, Toyonaka, Osaka 560\\ 
$^{2}$ Department of Applied Physics,  
University of Tokyo, Hongo, Tokyo 113}
\date{\today}
\maketitle

\begin{abstract}
Quantum mechanical motion 
of a particle in a periodic asymmetric potential is studied theoretically 
at zero temperature. 
It is shown based on semi-classical approximation that 
the tunneling probability from one local minimum to the next 
becomes asymmetric in the presence of weak oscillating field,
even though there is no macroscopic field gradient in average. 
Dissipation enhances this asymmetry,
and leads to a steady unidirectional current, 
resulting in a quantum ratchet system.  
\end{abstract}

\newcommand{\rmR}{{\rm R}}
\newcommand{\rmL}{{\rm L}}
\newcommand{\tilom}{\tilde{\omega}}

Ratchet is a mechanism with a periodic asymmetric potential 
where unbiased external perturbation leads to 
an unidirectional motion.
Such system is realized in a macroscopic scale 
by use of a pawl and a wheel having inclined teeth, 
in which the pawl prevents the wheel from rotating in either of the two 
directions, as is utilized in hand tools. 
Feynman in his famous lecture\cite{Feynman} discussed a ratchet system 
driven by the thermal fluctuation.
Here the existence of dissipation 
in addition to the fluctuating force is indispensable;
otherwise the pawl will keep bouncing after it hits the wheel 
and thus the wheel can turn in the other way.

Recently a Brownian particle in a periodic asymmetric potential 
subject to an additional external force having time correlation 
has been extensively investigated as a microscopic realization of 
ratchet\cite{AP92,Magnasco,Doering,Faucheus,Astumian,Zapata,HL97}.
Several cases of forces, e.g.,
a sinusoidally oscillating field\cite{Magnasco}, a colored noise
\cite{Magnasco} 
and a random telegraph noise\cite{Doering} have been studied.  
The time correlation in such noises breaks the detailed balance 
and thus can allow the asymmetric motion.\cite{Cu94,diode}  
These studies dealt with the case of overdamped limit.
Different mechanisms of a ratchet have also been pointed 
out\cite{Buttiker,Hondou}.
A case of space dependent diffusion constant was investigated in 
ref.\cite{Buttiker}.
In ref.\cite{Hondou},
a case of a chaotic noise was treated whose time correlation is asymmetric 
even though its invariant measure is uniform.
In these systems, an unidirectional current can emerge
even in the case of a symmetric potential.  
Microscopic ratchets have recently been fabricated 
by use of charged colloidal particles 
in an array of asymmetric electrodes 
with the electronic field turned on and off\cite{Rousselet} 
and by use of an array of small asymmetric antidots\cite{Lorke}.
The fluctuation driven ratchet has been studied 
also as a model to describe molecular motors\cite{VO90,Astumian}.

One might then wonder if such ratchet is possible in quantum system. 
Very recent study indicates that the answer is yes\cite{Reimann,Yukawa}. 
Reimann et.al. investigated the quantum mechanical motion of the particle 
under the asymmetric periodic potential and ac external field 
interacting with the heat-bath degrees of freedom\cite{Reimann}.
The asymmetric flow in the low temperature region is due to the difference 
between the two tunneling rates to the neighboring minima.
The current is in the opposite direction to that in the classical regime.
The present authors have investigated numerically 
the current of a quantum particle in the asymmetric potential 
with the external force of on-off type 
and interacting with heat bath based on a tight binding model\cite{Yukawa}. 
The current found there turned out to be due to 
the synchronous oscillation in the on-potential period, 
which is stabilized by the interaction with the heat bath.
Interestingly the result indicated that the direction of the current can be
controlled 
by changing the parameters of the external force, 
in contrast to the classical case.
So the induced current there is a manifestation 
of a new quantum mechanical phenomenon 
resulting from the interaction with the heat bath. 

In this letter we present an analytical calculation of the asymmetric 
transport 
in quantum system based on the path integral formalism.  
We consider a particle with mass $m$ in one-dimension, 
whose position is described by a coordinate $x$,
interacting with heat bath at zero temperature.
The potential, $U(x)$ is of saw-tooth shape with the period $a$, 
as shown in Fig. \ref{FIGpot}, 
a unit of which being written as 
\begin{equation}
U(x)=\alpha (a-x) \;\;\;\; (0\leq x \leq a ) , 
\end{equation}
where $\alpha$ is the slope of the potential. 
We apply an oscillating field of the amplitude $\epsilon_{0}$ 
and the frequency $\omega$. 
The Lagrangian of the system in the absence of the heat bath
is expressed as follows:
\begin{equation}
L_{0}=\frac{m}{2}\left(\frac{dx}{dt}\right)^{2} -U(x) -\epsilon_{0}x\cos
\omega t.
\label{L0}
\end{equation}
As a heat bath we consider the case of 
the Ohmic dissipation with the friction coefficient $\eta$.
The total action is then written as\cite{CL} 
$S=S^{0}+S'$, where $S^{0}\equiv \int dt L_{0}$ and
\begin{equation}
S' = -\frac{\eta}{2\pi} \int dt \int dt' 
\frac{(x(t)-x(t'))^{2}}{(t-t')^{2}} . \label{Sp}
\end{equation}

We first study the tunneling of the particle initially at $x=0$ 
with zero energy into the right minimum at $x=a$ 
based on the semiclassical approximation.
The external field and the interaction with the heat bath 
are assumed to be weak, 
so that we treat them perturbatively up to the linear order of 
$\epsilon_{0}$ and $\eta$.
We consider the incoherent tunneling regime, that is,  
energy dissipates sufficiently during the real time motion after each 
tunneling and then the particle returns to the zero energy state again. 
The corresponding condition for $\eta$ is given later. 
The semiclassical expression of the tunneling rate to the right direction 
is written as\cite{Fisher,Ivlev}
\begin{equation}
\Gamma_{\rm R}\sim \omega_{0} e^{-S_{\rm R}/\hbar}, \label{GamR} 
\end{equation}
where $\omega_{0}$ is the frequency of the zero point oscillation 
around $x=0$.
The exponent $S_{\rm R}$ is the action along 
$x_{\rm R}(\tau)$,
which is the solution of the classical equation of motion
in the imaginary time, $\tau\equiv it$, 
connecting $x=0$ and $x=a$.
To the linear order of $\eta$ and $\epsilon_{0}$,
where $x_{\rm R}(\tau)$ is the solution in the absence of $S'$,
$S_{\rm R}$ is expressed as
\begin{eqnarray}
S_{\rmR} & = & \int_{-\tau_{\rmR}}^{\tau_{\rmR}} d\tau 
\left[ \frac{m}{2}\left( \frac{dx_{\rmR}}{d\tau} \right)^{2} 
+U(x_{\rmR}) -\epsilon x_{\rmR}
\cosh \omega\tau \right] \nonumber\\ && +\frac{\eta}{2\pi} 
\frac{1}{2}\left[
\int_{-\tau_{\rmR}}^{\tau_{\rmR}} d\tau
\int_{-\tau_{\rmR}}^{\tau_{\rmR}} d\tau' + \int_{0}^{2\tau_{\rmR}} 
d\tau
\int_{0}^{2\tau_{\rmR}} d\tau' \right]
\frac{(x_{\rmR}(\tau)-x_{\rmR}(\tau'))^{2}}{(\tau-\tau')^{2}} 
\nonumber \\ & \equiv & S_{\rmR}^{0}+ S_{\rmR}'.
\label{SR}
\end{eqnarray}
Here $S_{\rmR}^{0}$ and $S_{\rmR}'$ denotes the non-dissipative 
and dissipative part, respectively. 
The origin of the imaginary time, $\tau=0$, is chosen as the time 
when the particle emerge from the barrier near $x\sim a$
and $\tau_{\rmR}$ is the imaginary time 
needed to go through the barrier, which will be determined later.
The amplitude of the field here is
$\epsilon\equiv\epsilon_{0} \cos\omega t_{0}$, $t_{0}$ 
being the real time 
when the particle entered the barrier.
The time integration in the dissipative part, $S_{\rmR}'$, is averaged 
over the contribution from the two time intervals, 
$-\tau_{\rmR} \leq \tau \leq \tau_{\rmR}$ and $ 0 
\leq \tau \leq 2\tau_{\rmR}$, 
because of the periodic behavior of the classical solution.

The classical path $x_{\rmR}$ is determined 
by the imaginary time equation of motion
derived form $S_{\rmR}^{0}$,
\begin{equation}
m\frac{d^{2}x_{\rmR}}{d\tau^{2}} -\alpha -\epsilon\cosh\omega \tau=0 . 
\label{eqR}
\end{equation}
One of the boundary conditions arises 
from the requirement that the particle is at the turning point at $\tau=0$, 
i.e., $dx_{\rmR}(\tau=0)/d\tau=0$.  
Other conditions are $x_{\rmR}(-\tau_{\rmR})=0$ 
and $(dx_{\rmR}(-\tau_{\rmR})/d\tau)=\sqrt{2\alpha a/m}$, 
where the initial velocity at $\tau=-\tau_{\rmR}+0$ is due to the jump of
the potential energy, $\alpha a$, at $x=0$.
These conditions also determine $\tau_{\rmR}$.  
It is easy to see that such classical solution exists only 
when $\epsilon \geq 0$,
namely the particle can tunnel to the right direction 
at the instant $t_{0}$ such that $\cos \omega t_{0}\geq 0$.
Thus $\epsilon$ below should be read as the average over $t_{0}$ 
(denoted by brackets), $\epsilon \simeq  \epsilon_{0}\langle |\cos\omega 
t_{0}|\rangle =2\epsilon_{0}/\pi$.
The solution for the time interval 
$-\tau_{\rmR} \leq \tau \leq\tau_{\rmR}$ is then given as
\begin{equation}
x_{\rmR}(\tau) = 
a\left[1- \left(\frac{\tau}{\tau_{0}}\right)^{2} -\frac{2\epsilon}
{\alpha\tilom^{2}} 
(\cosh\omega \tau-1) \right. \left.
-\frac{2\epsilon}{\alpha\tilom^{2}} 
(\tilom\sinh\tilom-\cosh\tilom+1) \right]+O(\epsilon^{2}),
\label{xRsol}
\end{equation}
where $\tau_{0}\equiv \sqrt{2ma/\alpha}$ is the time 
when the particle reach the turning point in the absence of the field 
and $\tilom\equiv\omega\tau_{0}$.  
The exit point,
$x_{\rmR}(0)=a[1-(2\epsilon\cosh\tilom/\alpha\tilom)]$, 
is closer to the origin than that in the case of $\epsilon=0$.
The traversal time, $\tau_{\rmR}$, is obtained as 
\begin{equation}
\tau_{\rmR}=\tau_{0}\left[ 1-
\frac{\epsilon}{\alpha}\frac{\sinh\tilom}{\tilom} \right]+O(\epsilon^{2}). 
\label{tauR}
\end{equation}
It is seen that this time is reduced by the oscillating field, 
because the field increases the effective force acting on the particle 
($\propto (\alpha+\epsilon)$ at $\tau=0$, see Eq. (\ref{eqR})).  
By use of this solution the non-dissipative part of the action is 
calculated as 
\begin{equation}
S_{\rmR}^{0}=\frac{4}{3}\alpha \tau_{0}a -4\epsilon\tau_{0} 
a \frac{1}{\tilom^{3}} 
(\tilom\cosh\tilom-\sinh\tilom). \label{sR0val}
\end{equation}
Thus the tunneling rate without dissipation 
$\sim \omega_{0} e^{-S_{\rmR}^{0}}$
is enhanced by the oscillating field, 
and the effect grows exponentially for large 
$\omega$\cite{Fisher,Ivlev}.  
This enhancement is due to the excitation to higher energy levels 
by the field prior to the tunneling.  
It is seen from Eq.(\ref{sR0val}) 
that the perturbative treatment of $\epsilon$ is justified 
if $\epsilon e^{\omega\tau_{0}}/\alpha \ll 1$.  
The dissipative part of the action is]
\begin{eqnarray}
S_{\rmR}' & = & -8\frac{\epsilon}{\alpha}\frac{\eta a^{2}}{\tilom^{2}} 
\left[ \frac{8}{3}(4-5\ln2)\tilom\sinh\tilom \right. 
\nonumber\\ & & 
\left. +\int_{0}^{1}dx \int_{0}^{1}dy 
\frac{x^{2}+y^{2}}{x^{2}-y^{2}} \left\{ \cosh\tilom x -\cosh\tilom y
+\frac{x+y-2}{x+y}(\cosh\tilom (1-x) -\cosh\tilom (1-y) ) \right\} \right]. 
\label{sRpval}
\end{eqnarray}

The tunneling probability to the left direction is calculated similarly.  
In this case, the classical path exists for $\epsilon \leq 0$.
It is obtained for $-\tau_{\rmL}\leq\tau\leq \tau_{\rmL}$ as 
\begin{equation} x_{\rmL}(\tau)=-a\left[
\frac{(|\tau|-\tau_{0})^{2}}{\tau_{0}^{2}}-2\frac{|\epsilon|}
{\alpha\tilom^ {2}}
\left(\cosh\omega\tau -1 -\frac{|\tau|}{\tau_{0}}
(\cosh\tilom-1) \right)\right]+O(\epsilon^{2}),
\label{xL}
\end{equation}
where the time $\tau_{\rmL}$ is given as 
\begin{equation}
\tau_{\rmL}=\tau_{0}\left[ 1+\frac{2|\epsilon|}{\alpha\tilom^{2}} 
(\tilom\sinh\tilom-\cosh\tilom+1)\right]+O(\epsilon^{2}). \label{tL} 
\end{equation}
It is seen that longer time is needed here for the particle
to reach the exit point in the presence of the external field, 
than
the case of tunneling to the right direction (compare with Eq. (\ref{tauR})).  
Essentially this difference in the traversal time leads to the difference 
between the tunneling rates to the two directions.  
By use of this solution the non-dissipative 
and the dissipative part of the action is calculated as
\begin{equation}
S_{\rmL}^{0}=\frac{4}{3}\alpha \tau_{0}a -4|\epsilon|\tau_{0} 
a \frac{1}{\tilom^{3}} (\sinh\tilom-\tilom), 
\label{sL0val} \end{equation}
and
\begin{eqnarray}
S_{\rmL}' & = & -8\frac{|\epsilon|}{\alpha}\frac{\eta a^{2}}
{\tilom^{2}} 
\left[ 4(1-\ln2)(\cosh\tilom-1) \right. 
\nonumber\\ & & \left. +\int_{0}^{1}dx \int_{0}^{1}dy 
\frac{x^{2}+y^{2}}{x^{2}-y^{2}} 
\left\{ \cosh\tilom (1-x) -\cosh\tilom (1-y) 
+\frac{x+y-2}{x+y}(\cosh\tilom x -\cosh\tilom y ) \right\} \right], 
\label{sLpval}
\end{eqnarray}
respectively.

Therefore the net current $J$, defined by the deference 
of the tunneling rate, is obtained as 
\begin{equation} 
J\equiv \Gamma_{\rm R}-\Gamma_{\rm L} \simeq 
\omega_{0}e^{-{4}\alpha \tau_{0}a/3\hbar}4|\epsilon|\tau_{0} 
a (I(\tilom)+\eta a^{2} K(\tilom))
+O(\epsilon^{2},\eta^{2}),
\label{Jdef}
\end{equation}
where
\begin{equation}
I(\tilom)\equiv
\frac{1}{\tilom^{3}} (\tilom\cosh\tilom-2\sinh\tilom+\tilom) ,
\nonumber\\ 
\end{equation}
and
\begin{eqnarray}
K(\tilom)&\equiv& \frac{1}{3} \left[
(51-64\ln2)\frac{\sinh\tilom}{\tilom} +
(31-36\ln2)\frac{\cosh\tilom-1}{\tilom^{2}} \right] 
\nonumber\\ & & -\frac{2}{\tilom^{2}}
\int_{0}^{1}dx \int_{0}^{1}dy \frac{(x^{2}+y^{2})(x+y-1)}{(x-y)(x+y)^{2}} 
\left( \cosh\tilom x -\cosh\tilom y 
+\cosh\tilom (1-x) -\cosh\tilom (1-y) ) \right)
.
\label{IJdef}
\end{eqnarray}
These functions are positive and
increase with $\tilom$, as plotted in Fig. \ref{FIGIJ}.  
Thus the current flow is in the right direction, 
which is consistent with the result
by Reimann et.al. \cite{Reimann}.
Since $K$ is much larger than $I$, 
the asymmetric current increases rapidly as dissipation becomes stronger.  
It is interesting to compare the result, Eq. (\ref{Jdef}), 
with that of classical ratchets
\cite{Magnasco,Doering,Faucheus,Hondou,Astumian} at low temperature, 
where the current vanishes if the external perturbation described by 
$|\epsilon|$ is smaller than a finite threshold value. 

So far we have looked into the asymmetry in a single tunneling process.  
This asymmetry leads to a net current 
in the incoherent tunneling regime we are assuming here.
In this regime the particle dissipates all the energy after each tunneling. 
Otherwise the particle gain and gain the energy 
from the oscillating field 
and then eventually will not feel the asymmetric potential 
when its energy exceeds the barrier height.
The condition of $\eta$ to realize this situation is obtained in a following way.
The increase of the energy of the particle 
after a single tunneling to the right direction is evaluated as 
\begin{equation} \Delta E \equiv U(x_{\rmR}(0))=2\epsilon 
a\frac{\sinh\tilom}{\tilom}. \label{delE} \end{equation}
In the present case of Ohmic dissipation the friction force 
is written as $F_{\rm dis}=-\eta dx/dt$ in the classical limit,
and then the rate of energy dissipation per time, $\eta_{\rm E}$ , 
is given as $\eta_{\rm E}\sim \eta\langle (dx/dt)^{2}\rangle $ 
(the bracket denotes the average).
This quantity, 
being proportional to the kinetic energy after the tunneling, 
is then estimated as
$\eta_{\rm E}\sim \eta\Delta E/m$.
The time interval between two successive tunnelings is about 
$\Gamma_{\rmR}^{-1}\simeq \omega_{0}^{-1}e^{4\alpha\tau a/3\hbar} 
\equiv \Gamma_{0}^{-1}$.
Thus the energy dissipated in this interval is 
\begin{equation}
\Delta E_{\rm dis}=\eta_{\em E}
\Gamma_{0}^{-1} %
\sim \frac{\eta\Delta E}{m\omega_{0}}e^{4\alpha\tau a/3\hbar }. 
\label{delEdis}
\end{equation}
The tunneling becomes incoherent if this is larger than $\Delta E$, 
{\it i.e.}, 
\begin{equation}
\eta \gsim m\omega_{0}e^{-4\alpha\tau a/3\hbar }. 
\label{etacondition} \end{equation}
When this condition is satisfied, a steady net current emerges.
From an experimental point of view, it may be interesting also
to observe a transitory current in the case of no or very weak dissipation, 
where the above condition is not satisfied.
Such current will survive for a time about
$(\alpha a / \Delta E \Gamma_{0}) \sim
(\alpha/2\epsilon)(\tilom/\sinh\tilom)e^{4\alpha\tau a/3\hbar}
\omega_{0}^{-1}$.

The tunneling rate and hence the resulting asymmetric current found 
here will show a resonant behavior
due to the small oscillation around the minimum 
before the tunneling\cite{Fisher,Ivlev}, 
which is not properly taken into account in the present calculation.
Then the current will be enhanced at the frequency 
$\omega \simeq n \omega_{0}$, with $n$ being integer.

Our study suggests the interesting
possibility of observation ratchets, 
where instead of the heat bath a periodic observation of the system by, 
{\it e.g.}, 
a laser pulse, is used to realize an incoherent tunneling regime.  
The wave function of the particle initially at the origin of the potential 
$U$ will spread asymmetrically with time
under the oscillating field.  
After a time of about $\Gamma_{0}^{-1}$ 
the particle has a good chance of being at either of the two minima, 
$x=\pm a$, 
but with a larger probability of being at $x=a$.
If we observe the system at that moment the wave function will shrink 
there with biggest probability.
Successive observation will then make the particle to diffuse 
to the right direction in average.

The realizations of the quantum ratchet will be attained 
in a system of asymmetric SQUID.
As demonstrated by Zapata {\it et al.}\cite{Zapata}, 
the phase variable can experience a periodic sawtooth potential 
in a SQUID containing three identical Josephson junctions
with resistance $R$ and capacitance $C$, two being on one of the arms of 
the ring (we call L) and one on the other (R).
The equation of motion for the phase $\varphi$ of a L--junction 
in this case is \begin{equation}
\frac{\hbar C}{e}\ddot{\varphi}+\frac{\hbar}{eR}\dot{\varphi} +J_{L}
\sin\frac{\varphi}{2}
+J_{R}\sin({\varphi}+\varphi_{\rm ext}) =I(t), 
\label{eqSQUID}
\end{equation}
where $J_{L}$ ($J_{R}$) is the critical current of L(R)--junction
and $I(t)$ is the total current through the SQUID. 
$\varphi_{\rm ext}\equiv 2\pi\Phi_{\rm ext}/\Phi_{0}$ 
is the phase due to an external flux $\Phi_{\rm ext}$ ($\Phi_{0}$ 
being the flux quantum).  
By choosing values of $J_{R}/J_{L}$ and $\varphi_{\rm ext}$, 
a sawtooth potential can be realized.
From our analysis, if ac current $I(t)=I_{0}\cos\omega t$ is 
applied, finite voltage $V=\hbar \langle \dot\varphi \rangle /2e$ appears 
due to the directional drift of $\varphi$.

Recently an array of antidots of triangular shape with dimension 
of about 2--300nm has been fabricated on GaAs
substrate, where electrons passing 
through the array feel an asymmetric potential\cite{Lorke}.
Applying a far-infrared irradiation of 119$\mu$m wavelength, 
a finite photovoltage has been observed, 
which indicates an asymmetric classical transport of the electrons.
In such antidot systems, 
the density of the electrons can easily be controlled, 
and thus they would be suitable for the realization of the quantum ratchets.  
A new electronic device might be possible 
by use of such systems of small semi-conductors, 
where the external perturbation leads to a density gradient of electrons.

In the system we considered here 
the current at zero temperature is always in the right direction. 
This is in contrast to the case of the tight binding model 
subject to an on-off type potential considered previously by the present
authors\cite{Yukawa}. 
There the current changes its direction
with parameters of the external field.
This difference might be due to the difference 
of the type of the external perturbation.
In the present system,  
an oscillating field leads to an asymmetric probability 
of a single tunneling event, while in the system studied 
previously, the particle oscillates in the well 
when the potential is on 
and its wave function begins to spread 
when the potential is turned off.
The oscillation in the on-potential period is averaged out 
and does not contribute to the net flow.
The net flow results form the motion in the off potential period.
The direction of the net flow is then determined by 
the phase of the oscillation in the on-potential period 
at the instant the potential is switched off.
Such a oscillation is stabilized  by the interaction with the heat bath, 
the phase mentioned above becomes constant, 
and then the net flow becomes finite.

In summary we have shown at zero temperature 
that an asymmetric flow of a quantum dissipative particle 
arises in the presence of asymmetric periodic potential 
and an oscillating field.
Effects of dissipation are twofold:
to enhance the asymmetry of the rate of the individual tunneling,
and to lead to a steady unidirectional current.
This quantum ratchet system exhibits a current 
linear in the amplitude of the field, 
$\epsilon_{0}$, as the lowest contribution, 
in contrast to the classical systems 
at low temperature, where there is a finite threshold value of 
$\epsilon_{0}$.

We thank 
H. Akera, O. Matsuda, A. Kawabata, S. Miyashita, K. Saito and T. Nagao
for useful comments.
G.T. thanks The Murata Science Foundation for financial support. 
H. M. thanks H. Fukuyama for sending him a copy of ref. \cite{Lorke}.
This work is partly supported by Grants-in-Aid from the Ministry of Education, 
Science, Sports and Culture.

\begin{figure}
\caption{Asymmetric potential barrier. \label{FIGpot}} 
\caption{Behaviors of $I(\tilom)$ 
and $K(\tilom)$ as a function of the dimensionless frequency, 
$\tilom\equiv \omega\tau_{0}$. \label{FIGIJ}} \end{figure} 

\begin{thebibliography}{99}
\bibitem{Feynman} {\it The Feynman Lectures on Physics}, 
Vol. I, Chap. 46 (Addison-Wesley, 1963).
\bibitem{AP92} A. Ajdari and J. Prost: C. R. Acad. Sci. Paris, {\bf 315} 
(1992) 1635.
\bibitem{Magnasco} M. O. Magnasco: Phys. Rev. Lett. 
\bibitem{Doering}
C. R. Doering, W. Horsthemke and J. Riordan: 
Phys. Rev. Lett. {\bf 72} (1994) 2984.
\bibitem{Faucheus}
L. P. Faucheus, L. S. Bourdieu, P. D. Kaplan and A. J. Libchaber: 
Phys. Rev. Lett. {\bf 74} (1995) 1504.
\bibitem{Astumian} R. D. Astumian and M. Bier: Phys. Rev. Lett. 
{\bf 72}  (1994) 1766.
\bibitem{Zapata} I. Zapata, R. Bartussek, F. Sols and P. H\"{a}nggi: 
Phys. Rev. Lett. {\bf 77} (1996) 2292. 
\bibitem{HL97} T. Harms and R. Lipowsky: Phys. Rev. Lett. {\bf 79}
  (1997) 2895. 
{\bf 71} (1993) 1477. 
\bibitem{Cu94} P. M. P. Curie: J. de Phys. III. {\bf 3} (1894) 393.
\bibitem{diode}
The mechanism of the asymmetric flow in these ratchet systems 
are different from that of the rectification of diodes, 
which is due to the potential difference across the junction 
introduced from the start and the rearrangement of electrons upon bias. 
\bibitem{Buttiker} M. B\"{u}ttiker: Z. Phys. B {\bf 68} (1987) 161.
\bibitem{Hondou} T. Hondou and Y. Sawada: Phys. Rev. Lett. 
{\bf 75} (1995) 3269.
\bibitem{Rousselet}J. Rousselet, L. Salome, A. Ajdari and J. Prost: 
Nature {\bf 370} (1994) 446.
\bibitem{Lorke} A. Lorke, S. Wimmer, B. Jager, J. P. Kotthaus, 
W. Wegscheider and M. Bichler, prerint.
\bibitem{VO90} R. D. Vale and F. Oosawa: Adv. Biophys. {\bf 26} 
(1990) 97.
\bibitem{Reimann} P. Reimann, M. Grifoni and P. H\"{a}nggi: 
Phys. Rev. Lett. {\bf 79} (1997) 10.
\bibitem{Yukawa} S. Yukawa, M. Kikuchi, G. Tatara and H. Matsukawa, 
J. Phys. Soc. Jpn., {\bf 66} (1997) 2953.
\bibitem{CL} A. O. Caldeira and A. J. Leggett: 
Phys. Rev. Lett. {\bf 46} (1981) 211;
A. O. Caldeira and A. J. Leggett: Ann. Phys. {\bf 149} (1983) 374. 
\bibitem{Fisher} M. P. A. Fisher: Phys. Rev. {\bf B37} (1988) 75. 
\bibitem{Ivlev} B. I. Ivlev and V. I. Mel'nikov: Sov. Phys. JETP 
{\bf 62} (1985) 1298; 
in {\it Quantum Tunneling in Condensed Media}, 
eds. Y. Kagan and A. J. Leggett, Chap. 5 (North-Holland, 1992). 
\end{thebibliography}
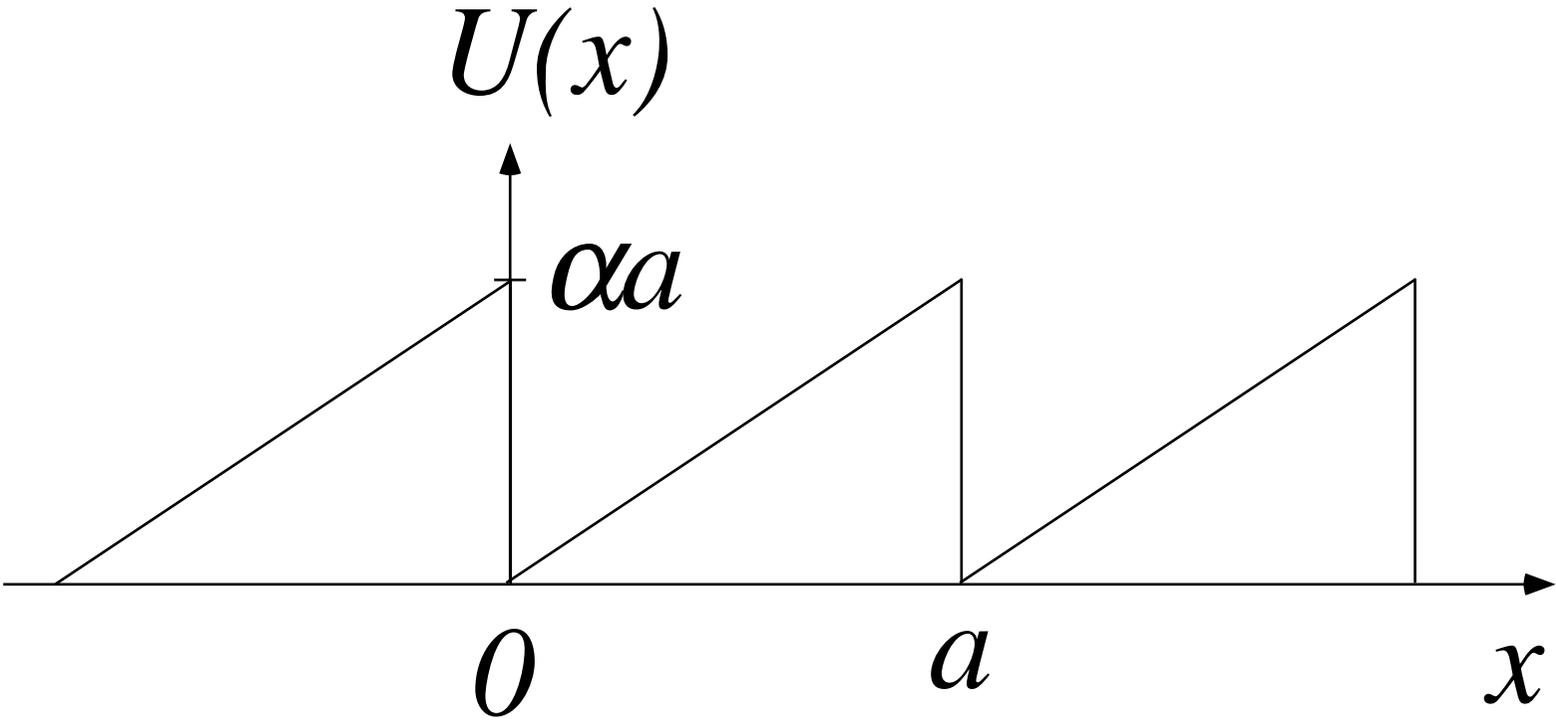
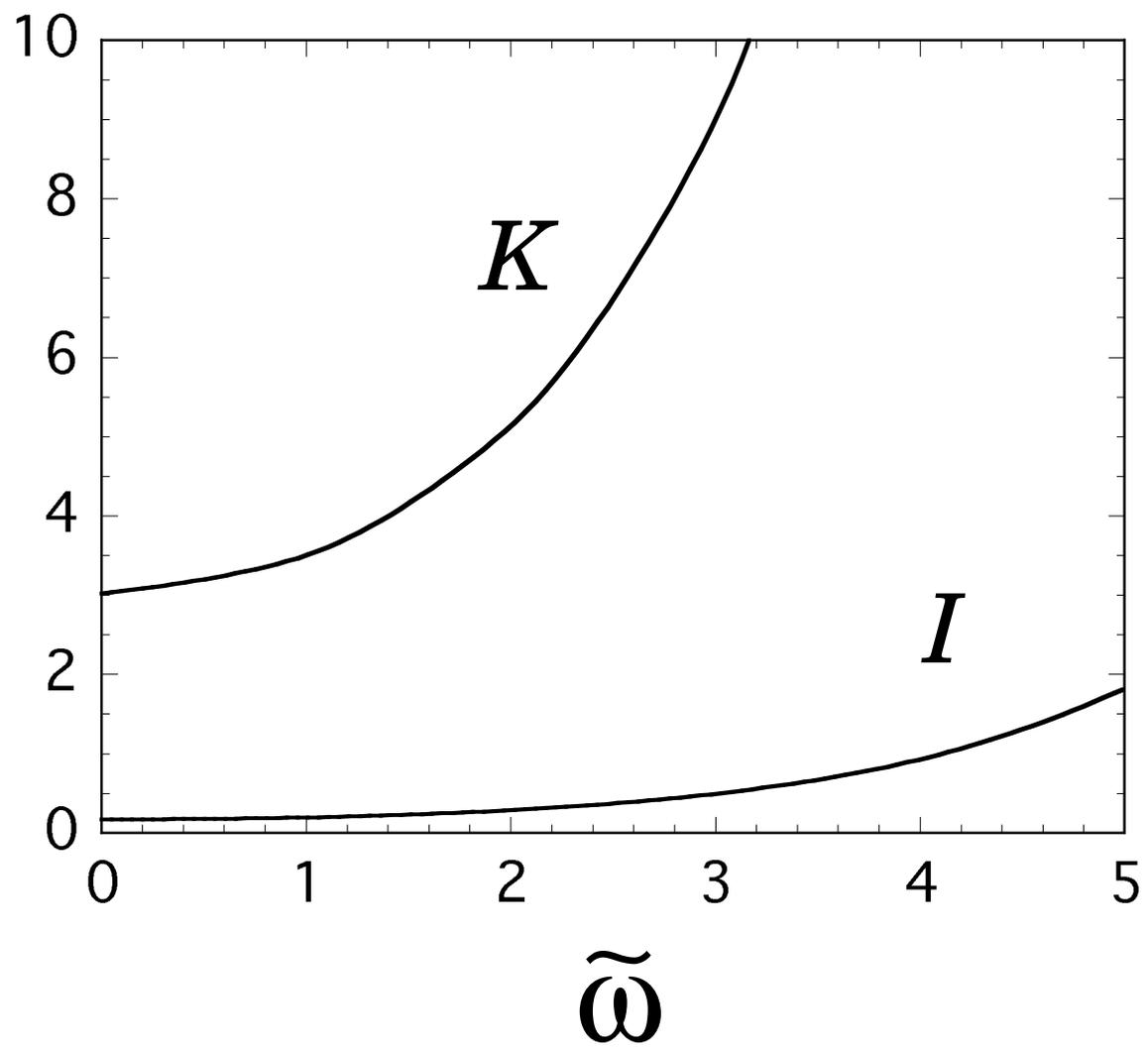
\end{document}